\documentclass[reprint,aps,prl,twocolumn,showpacs,preprintnumbers,amsmath,amssymb,amsfonts]{revtex4-1}

\usepackage{graphicx}
 \usepackage{epstopdf}
 \usepackage{color}
\usepackage{siunitx}
\usepackage{natbib}
\begin{document}

\title{Spectral modification of shock accelerated ions using hydrodynamically shaped gas target}

\author{O.~Tresca$^{1}$}
\author{N.~P.~Dover$^{2}$}
\author{N.~Cook$^{3}$}
\author{C.~Maharjan$^{3}$}
\author{M.~N.~Polyanskiy$^{1}$}
\author{Z.~Najmudin$^{2}$}
\author{P.~Shkolnikov$^{2}$}
\author{I.~Pogorelsky$^{1}$}

\affiliation{$^{1}$ Accelerator Test Facility, Brookhaven National Laboratory, Upton, New York 11973, USA}
\affiliation{$^{2}$ John Adams Institute for Accelerator Science, Blackett Laboratory, Imperial College London, SW7 2BZ, United Kingdom}
\affiliation{$^{3}$ Stony Brook University, Stony Brook, New York 11794, USA}

\date{\today}

\begin{abstract}
We report on reproducible shock acceleration from irradiation of a $\lambda = 10$\,$\mu$m CO$_2$ laser on optically shaped H$_2$ and He gas targets. A low energy laser prepulse ($I\lesssim10^{14}\, {\rm Wcm^{-2}}$) was used to drive a blast wave inside the gas target, creating a steepened, variable density gradient. This was followed, after 25\,ns, by a high intensity laser pulse ($I>10^{16}\, {\rm Wcm^{-2}}$) that produces an electrostatic collisionless shock. Upstream ions were accelerated for a narrow range of prepulse energies ($> 110$ mJ \& $< 220$mJ). For long density gradients ($\gtrsim 40 \mu$m), broadband beams of He$^+$ and H$^+$ were routinely produced, whilst for shorter gradients ($\lesssim 20 \mu$m), quasimonoenergetic acceleration of proton was observed. These measurements indicate that the properties of the accelerating shock and the resultant ion energy distribution, in particular the  production of narrow energy spread beams, is highly dependent on the plasma density profile. These findings are corroborated by 2D PIC simulations.
 \end{abstract}

\maketitle

The light and thermal pressure associated with intense lasers causes compression and heating when incident on an overdense plasma. This can generate a piston that launches electrostatic collisionless shocks into the plasma \cite{denavit1992}. Upstream ions can be reflected off the shock creating an ion population accelerated to twice the velocity of the driving shock \cite{silva2004, wei2004, chen2007,fiuza2012}. This mechanism has been demonstrated in recent experiments using intense CO$_2$ lasers interacting with gas jets \cite{palmer2011,haberberger2012} where multi-MeV proton beams with energy spread smaller than 4\% have been reported.  

Gas jets have also been proposed as a source of high purity, high-Z ion beams. This is in contrast with the multiple species beams generated from solid targets via sheath acceleration \cite{clark2000, snavely2000}. For solid targets, protons from surface contaminants are preferentially accelerated due to their higher charge-to-mass ratio, making the production of impurity free high-Z beams challenging. Generating high purity helium ion beams from gas jets would be of interest for nuclear and medical physics applications \cite{daido2012}. 

Laser driven longitudinal \cite{willingale2006} and transverse \cite{wei2004,sylla2012, lifschitz2014} acceleration of helium ions from laser irradiation of gas jets has been observed experimentally. However, these previous studies were conducted in underdense plasmas requiring laser intensities $> 10^{20}$\,Wcm$^{-2}$. Collisionless shock acceleration, on the other hand, allows the production of directed beams at lower intensities with sizeable number \cite{palmer2011}, and with favourable intensity scaling \cite{fiuza2012}.

It is thought that the generation and subsequent properties of shock accelerated ion beams are highly dependent on the initial plasma density distribution \cite{dover2014, fiuza2012}. Gas jet targets used for laser-plasma interactions typically have initial density scale lengths $l>100$\,$\mu$m, which is unsuitable for efficient shock generation. In previous experiments using intense CO$_2$ lasers, this profile was modified by the pulse train inherent to the laser system \cite{palmer2011,haberberger2012}, which is difficult to control and reproduce. Recent breakthroughs in CO$_2$ laser technology have made single pulses of intense radiation \cite{polyanskiy2011} possible. This allows reproducible interaction of the laser with the target. However it requires a different method to modify the density profiles. One method investigated previously is the use of optically generated hydrodynamic blast waves via laser solid interaction \cite{hsieh2006, kaganovich2011}.

In this paper, we demonstrate the use of a controlled low energy prepulse focused in dense gas to generate blast waves that hydrodynamically shape the gas targets to make them suitable for ion acceleration. Interaction of the high intensity laser pulse with the modified plasma density profile leads to reproducible shock acceleration. In particular, we present the first measurements of shock accelerated, $>1$\,MeV, helium ions. Using hydrogen, we also demonstrate that shock acceleration can lead to both broadband and quasi-monoenergetic ions beams depending on the steepness of the initial plasma density gradient The role of the density gradients is further investigated with PIC simulations.

The experiment was performed at the Accelerator Test Facility at the Brookhaven National Laboratory. A linearly polarized CO$_2$ laser beam, $\lambda = 10.3$\,$\mu$m, with energy up to 11\,J per pulses of 5\,ps full-width-half-maximum (FWHM), provided a peak power of 2.2\,TW. An $f/3$ parabolic mirror focussed the laser to a spot $w_0=65$\,$\mu$m, resulting in peak intensity $I=2.5\times10^{16}$\,Wcm$^{-2}$ ($a_0=1.4$). The laser was focussed at the center of a helium or hydrogen gas-jet from a 1\,mm diameter supersonic nozzle, $\sim$ 700\,$\mu$m above the the nozzle. At this position the initial gas density profile, along the laser axis, was near triangular with a linear gradient of $\sim$ 1\,mm up to a peak fully-ionized plasma density of $n_{e}\sim2 n_c$ \cite{najmudin2011}.    

The ion beam was characterized using a laser-axis Thomson parabola spectrometer, opening angle of $2\times 10^{-5}$\,sr, and detected with BC-408 polyvinyl-toluene scintillator coupled to an EMCCD camera. The ion-to-detected-photon yield was calibrated using the Tandem proton accelerator at Stony Brook University \cite{cook2014}. The calibration was scaled to He$^{+}$ by introducing a scaling factor of $0.3\pm0.1$ \cite{Blasse1994}.

A laser prepulse collinear to the drive pulse was generated to shape the gas. It arrived 25\,ns before the main laser pulse and contained a variable energy, $E_{pp}$, up to 1\,J in 5\,ps. The resulting plasma density distribution was characterized using optical probing. Two, time-adjustable, $\sim 10$\,ps long, $\lambda = 532$\,nm synchronized pulses were used for shadowgraphy and interferometry. They were set to allow characterization just before and after the drive pulse on a single shot.

\begin{figure}[t]
 \centering
 \includegraphics[width=0.45\textwidth]{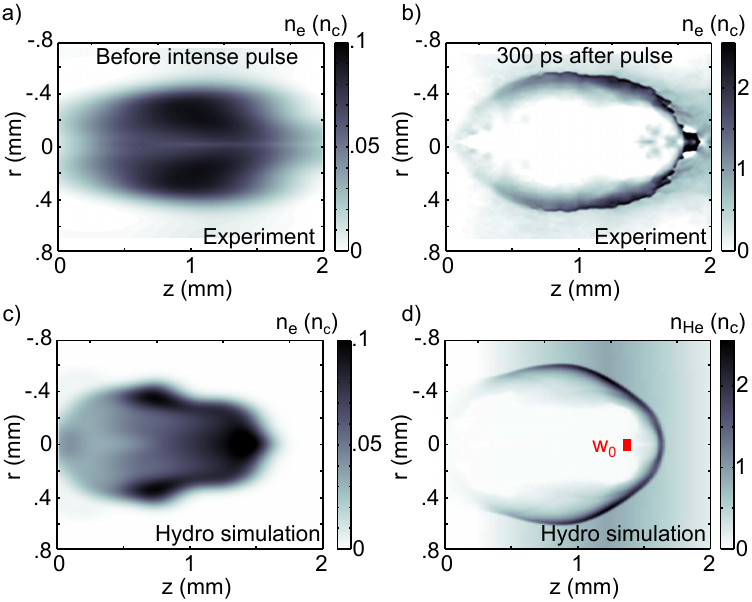} 
 \caption{ Electron plasma density from He targets, $n_{e}$ from interferometry: (a) immediately before, and (b) 300\,ps after the LPI, for $E_{pp} \approx 220$\,mJ. (c) $n_{e}$ and (d) total helium atomic density (irrespective of ionization), $n_{He}$ from hydro simulations after 25\,ns of expansion for 3\,mJ absorbed energy.}
\label{fig1}
 \end{figure}

Fig.\,\ref{fig1}a shows the experimentally observed plasma density map 100\,ps before the main laser plasma interaction (LPI) using He, similar images are found using H$_2$. For this shot, a prepulse $E_{pp} \approx 220$\,mJ was incident on helium gas of peak neutral density $n_{He} = 0.8\times10^{19}$\,cm$^{-3}$, corresponding to a fully ionized plasma density $n_e=1.6$\,$n_c$. The plasma density was obtained from interferometry assuming cylindrical symmetry around the laser axis, thus allowing Abel inversion. The optical path difference due to the unionized gas is small compared to that of the plasma and is therefore neglected. Plasma formation was observed over a large volume with a peak density significantly lower than the fully ionized density, and shows no structure. However, probing the plasma 300\,ps after the LPI revealed the characteristic high density shell of a blast wave (fig.\,\ref{fig1}b). The shell is observed to expand at velocity $< 10^5$\,ms$^{-1}$. Hence the dramatic change in the plasma appearance cannot be due to hydrodynamic motion originating from the main pulse interaction.

The effect of the prepulse was modelled using the {\sc flash} hydrodynamics solver in 2D cylindrical geometry. Helium was initialized following the experimental gas jet profile with $n_{He} = 0.8\times10^{19}$\,cm$^{-3}$. The prepulse energy was modelled by depositing thermal energy evenly into a cylinder of gas at the laser focus (length 600\,$\mu$m and radius 80\,$\mu$m) centered near the initial critical density surface; $n_{He} = \frac{1}{2} n_c$. Within 1\,ns, a collisional blast wave is formed and expands out into the cold gas \cite{zeldovich,raizer1966}. The blast wave forms a cavity with low density but high electron temperatures at the center, surrounded by a cold but high density shock moving outwards. 
\begin{figure}[b]  
\includegraphics[width=0.45\textwidth]{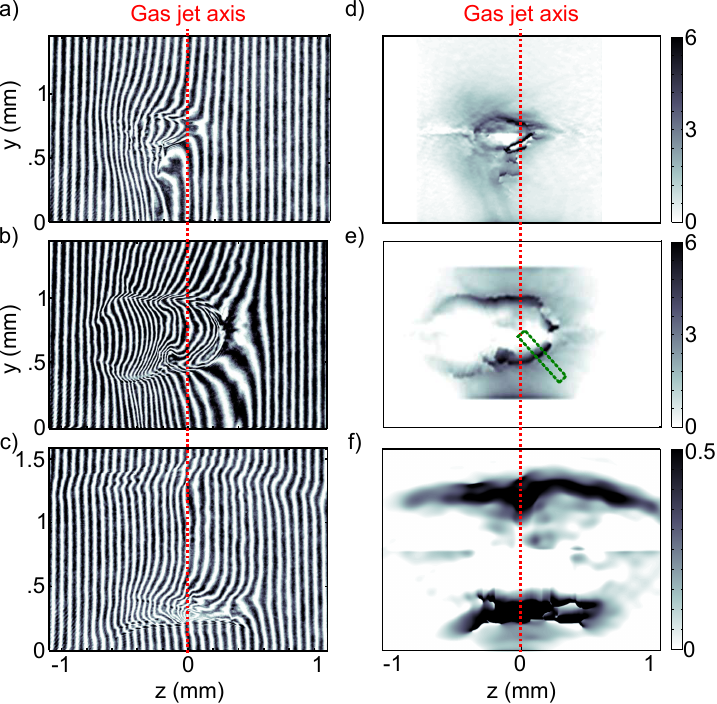}
\caption{ Interferograms 300\,ps after the LPI with He; (a) without prepulse, (b) with prepulse ideal for ion acceleration, $E_{pp} \approx 150$\,mJ, (c) and with prepulse too large for ion acceleration, $E_{pp} \approx 1.27$\,J. (d-f) corresponding $n_{e} /n_{cr}$.}
\label{fig2}
 \end{figure}
Fig.\,\ref{fig1}c shows the electron density at 25\,ns after the prepulse, calculated from the Saha equation, for an initial 3\,mJ of energy absorbed in the plasma. Only the center of the cavity is hot enough to be ionized and detectable by interferometry, in agreement with the experimental measurements prior to the LPI in fig.\,\ref{fig1}a. Inspection of the neutral helium density (fig.\,\ref{fig1}d) reveals the blast wave, which has a cavity wall of peak density given by that of a strong shock $(\gamma + 1) /(\gamma -1)n_i$, where $n_i$ is the ion density and $\gamma$, the ratio of specific heats (5/3 for He, 1.4 for H$_2$)\cite{zeldovich}.  Therefore, using hydrogen with comparable laser parameters gives a higher density enhancement as well as a reduced scale length \cite{sedov}.  Even though this neutral profile exists before the LPI, it is only due to ionization by the fast electrons generated during the main interaction that it can be directly observed. Once ionized, the blast wave provides the ideal small scale-length plasma in which to generate collisionless shock waves.
 
Interferograms for three different prepulse levels with helium gas are shown in fig.\,\ref{fig2}a-c, with corresponding electron densities in fig.\,\ref{fig2}d-f. The images are all taken 300\,ps after the LPI to reveal the full blast wave structure. Similar images are obtained with H$_2$. To account for the vertical density gradient in the initial gas density, azimuthal symmetry is assumed in each half-cylinder above and below the laser axis. These regions were processed individually by Abel inversion, with continuity assumed at their interface.

With no prepulse (fig.\,\ref{fig2} a \& d), a small cavity of diameter $\approx 100$\,$\mu$m forms around the laser focal position. The intense laser has self-focussed and channelled part way into the long density scale length ($l \approx 1$\,mm), coupling most of the energy into the plasma ramp. In this case, no forward accelerated ions were observed.  
Introducing a prepulse with energy $E_{pp} \approx 150$\,mJ (fig.\,\ref{fig2} b \& e) shows a significant difference in the plasma distribution. A blast wave has been generated, so that the intense pulse interacts with a profile with steeper density gradient, $\approx 100$\,$\mu$m, and higher peak density. Energetic ions were consistently generated in this regime. For further increase in $E_{pp}$  (fig.\,\ref{fig2} c \& f), the prepulse induced blast wave propagates deeper into the jet and ultimately through it, leaving only an underdense remnant along the laser axis. For this higher prepulse, no ion beam was observed. Probing in this case, showed little change to the shape of the plasma before and after the LPI, except for a doubling of the density in the walls, consistent with ionization from He$^{+}$ to He$^{2+}$.

\begin{figure}  
\includegraphics[width=0.45\textwidth]{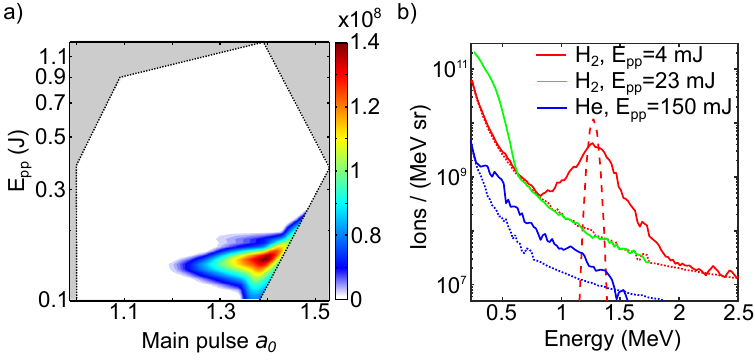}
\caption{(a) Average flux of accelerated helium ions (/sr)  $> 0.1$\,MeV as sampled by the spectrometer (color scale) as a function of prepulse energy, $E_{pp}$ and main pulse $a_{0}$. The white region indicates laser parameters tested experimentally. (b) Proton spectra for a H$_2$ target with $E_{pp} \approx 4$ and $ 23$ mJ, and He$^+$ spectrum for a He target with $E_{pp} \approx 150$\,mJ  (solid lines). The dotted lines are the detection thresholds, and the dashed red line is the deconvolved spectrum for $E_{pp} \approx 4$\,mJ.}
\label{fig3}
\end{figure}
 
Control of the helium density profile generated by the blast wave is essential for reproducible ion acceleration. Fig.\,\ref{fig3}a shows the flux of detected He$^{+}$ ions (color scale) as a function of main pulse normalized vector potential, $a_{0}$, and prepulse energy, $E_{pp}$.  Accelerated ions were only observed over a narrow range of prepulse energies, $110\,{\rm mJ} \lesssim E_{pp} \lesssim 220$\,mJ, showing the importance of optimizing the density profile as in fig.\,\ref{fig2} b \& d. Within this range of $E_{pp}$ and $a_o \approx 1.4$, the total flux of accelerated ions remains stable shot-to-shot at $ \approx 10^{8}$ ions/sr.

Fig.\,\ref{fig3}b shows a He$^{+}$ spectra obtained with $E_{pp} = 150\,$mJ and a main pulse $a_0=1.4$, resulting in a broad energy distribution up to 1.5\,MeV. This spectrum was typical for all the observed beams. Only He$^{+}$ ions are observed although the main pulse laser intensity exceeds the threshold value to double ionize helium via field ionization ($I\gtrsim 9 \times 10^{15}$\,Wcm$^{-2}$). The observed single charge state is the result of charge transfer as the ions traverse the plasma and un-ionized gas \cite{sylla2012,itoh1980}, and recombination as the accelerated ion beam co-moves with a low temperature electron cloud to the detector.

For fixed $n_i$ and absorbed energy $E_{abs}$, a blast wave in hydrogen will give a steeper density ramp than for helium due to the lower $\gamma$ \cite{sedov}.  Spectra from shots with H$_2$ for different prepulse levels $E_{pp}$ = 4\,mJ and 23\,mJ are also shown in fig.\,\ref{fig3}b.  Note that a significantly smaller prepulse was required for ion generation with hydrogen; observation of the relative blast wave size implies a significantly higher energy absorption.  For $E_{pp} = 4\,$mJ, a quasi-monoenergetic beam is observed with peak energy $\approx 1.2$ MeV and a deconvolved energy spread $\Delta E/E \approx 5$\,\%, compared with a broadband beam with energies up to $\approx 0.5$\,MeV for $E_{pp}=23$\,mJ.  As the radius for a spherical blast-wave $r_{bw} \propto (E_{abs}/n_i)^{1/5}$ \cite{zeldovich}, lower $E_{pp}$ results in less expansion, providing a steeper target density gradient. 
Not only does this technique allow the production of reproducible shock accelerated ions, but for sufficiently steep gradients spectral shaping is also achievable.  

 \begin{figure}[b]
 \includegraphics[width=0.45\textwidth]{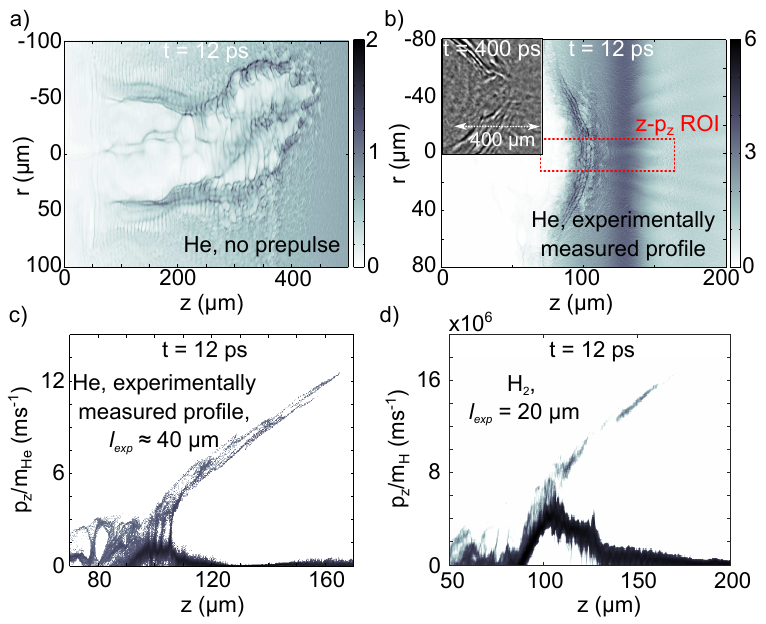}
 \caption{Results of PIC simulations. Ion density map 12\,ps after LPI for: (a) no prepulse, (b) experimental initial profile for $E_{pp}=150\,$mJ obtained from fig.\,\ref{fig2}e. (inset is optical shadowgraphy showing shock remnants) and (c)  $z$-$p_z$ phase space.  (d) The $z$-$p_z$ phase space for for hydrogen plasma with exponential ramp ($l_{exp}= 20$\,$\mu$m)}
\label{fig4}
 \end{figure}
 
The LPI was modelled using the 2D PIC code {\sc epoch}. The plasma was initialized with 30 particles per cell of cold ions (He$^{2+}$ or H$^+$) and electrons, in cells of size $\lambda/50$. The laser had $a_0 = 1.4$, $w_0 = 65$\,$\mu$m at focus, pulse length $\tau = 5$\,ps (FWHM), and was linearly polarized in the transverse plane. The initial density distribution was varied to investigate the effect of the optical density tailoring.

To model the LPI with no prepulse, the initial density profile was set to rise linearly from 0 to 2\,$n_c$ over 800\,$\mu$m with He$^{2+}$. The laser self-focuses and bores a channel in the plasma, shown in fig.~\ref{fig4}a.  All the laser energy is expended in the underdense region and no forward shock is formed, although lower velocity transverse electrostatic shocks are observed \cite{wei2004}.

The effect of the prepulse was simulated using a plasma density distribution extracted from the interferometry for $E_{pp} \approx 150\,$mJ (green box in fig.\,\ref{fig2}b). The profile was taken along a line radially off-set from the laser axis in order to reduce errors introduced by the Abel inversion.  In the blast-wave ramp the density is well approximated by $n=n_{max}e^{-(z_0-z)/l_{exp}}$, where $n_{max} \approx 6\,n_c$ is the peak density, $z_0$ is the position of the peak and $l_{exp} \approx 40$\,$\mu$m is the scale length.  In this simulation, the laser pulse shows a modest increase in intensity due to self-focussing in the ramp, but the short ramp length means the laser is not adversely absorbed. The pulse penetrates up to the critical surface where a combination of radiation and thermal pressure launches a collisionless shock (fig.\,\ref{fig4}b). Ions are reflected at the shock front to twice the shock speed $v_s$.  The result agrees well with optical shadowgraphy $300$\,ps after the LPI (inset of fig.\,\ref{fig4}b), which shows remnants of the accelerating collisionless shock emerging from the collisional prepulse induced blast wave.  

\begin{figure}  
\includegraphics[width=0.45\textwidth]{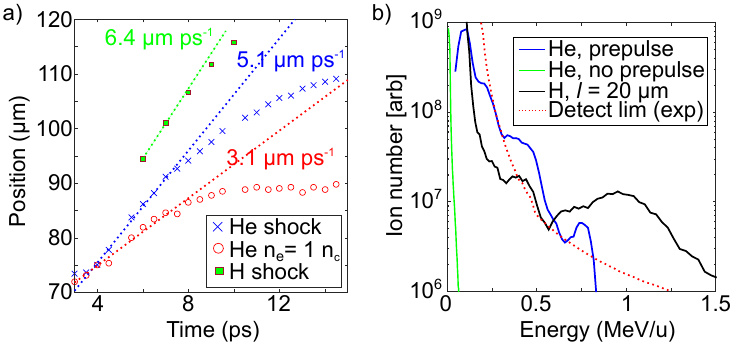}
\caption{ (a) Critical surface (red) and shock (blue) position with time for simulation shown in fig.\,\ref{fig4}b\&c, and shock position for hydrogen simulation in fig.\,\ref{fig4}d. (b) Comparison of He$^{2+}$ spectra for $800$\,$\mu$m linear ramp (no prepulse) (green),  prepulse fig.\,\ref{fig4}a\&b (blue), and for hydrogen with exponential profile, $l_{exp} = 20$\,$\mu$m (black).}
\label{fig5}
\end{figure}

The $z-p_z$ phase space (fig.\,\ref{fig4}c) demonstrates that the ions originate from the position of the shock, but also that the ions are reflected with a large energy spread.  Fig.\,\ref{fig5}a shows the position of the shock front and critical surface on the laser-axis as a function of time  for this simulation. The initial shock velocity $v_{sh} \approx 5\times10^6$\,ms$^{-1}$ exceeds the measured hole-boring velocity $v_{hb} = 3.1\times10^{6} \approx  \sqrt{I_L/n_im_ic}$ \cite{wilks1992,Robinson2009}.  Ions reflected off the shock at its peak velocity would gain $\sim 3.1$ MeV, in agreement with simulated maximum ion energies of $\sim 3.3$ MeV (fig.\,\ref{fig5}b).  Scaling the simulated ion spectra to the experimental spectra and imposing the experimental detection limit as in fig.\,\ref{fig3}b would result in a maximum detectable ion energy of  $\sim 2$ MeV, consistent with that measured. The energies in the case of a shaped target are significantly larger than for the no prepulse case.

The shock slows at the end of the laser drive while still reflecting ions, resulting in the large momentum spread in fig.\,\ref{fig4}b and the corresponding broadband axial ion spectrum in fig.\,\ref{fig5}b. The shock deceleration is mainly due to spherical expansion of the front. However, the experiment showed spectrally peaked proton beams were produced from sharper density gradients using hydrogen. Density profiles with an exponential ramp, and variable length scale, were used to reproduce the sharper profiles achievable with hydrogen. At the edge of the blast-wave, the density dropped to $n_{max}/6$, the density drop for a strong shock, imitating the experimental profile.  

A scale length $l_{exp} = 50$\,$\mu$m produces a broadband beam, as described for helium.  However, shortening the scale length to $l_{exp}=20$\,$\mu$m results in the shock breaking through the density discontinuity at the blast-wave front.  The sudden change in thermal pressure ahead of the front triggers the shock immediate dissipation, just at the end of the LPI (fig.\,\ref{fig5}a).  The axial $z-p_z$ phase space, fig.\,\ref{fig4}d, demonstrates the formation of a single bunch with velocity near $ 2 v_s \approx 13 \times 10^{6}$\,m/s.  Particle reflection stops and the peaked spectrum is maintained (fig.\,\ref{fig5}d).  This generation of peaked spectra differs from thermally driven shocks in isothermal plasmas \cite{haberberger2012,fiuza2012}, where reflection at a uniform speed persists over a longer time scale.  Further simulation with $l_{exp} < 10$\,$\mu$m with helium also demonstrated spectrally peaked beams.  Producing such gradients in helium would be experimentally achievable by shortening the time between the prepulse and the drive pulse, which was not possible with our experimental geometry. 

In this work, we have observed shock acceleration of protons and helium ions. The initial plasma density profile is found to be critical to achieve not only shock acceleration but also spectral control of the ion beam. Using hydrogen, we found that long density gradients lead to the production of broadband shock accelerated beams, while steeper gradients allow for the generation of quasi monoenergetic beams. Helium ions have also been accelerated using this scheme, demonstrating that it could be extended to other high Z gaseous species, providing a route towards a wide variety of easily replenishable high-repetition rate ion sources for future nuclear physics applications.

\begin{acknowledgments}
Work supported by the US DOE Grant DE-FG02-07ER41488, and UK EPSRC grant EP/K022415/1. {\sc flash} was developed by the DOE NNSA ASC and NSF-supported FCCS at the U.~Chicago. {\sc epoch} development was supported by EPSRC grants EP/G054940/1,  EP/G055165/1 and EP/G056803/1. Computing resources provided by Imperial College HPC services and NERSC (mp1401), supported by DOE Contract DE-AC02-05CH11231 and BNL/LDRD No.~12-032.
\end{acknowledgments}
 \end{document}